\documentclass[reprint,superscriptaddress,longbibliography,nofootinbib,nobibnotes,amsmath,amssymb,aps,prd,]{revtex4-2}
\pdfoutput=1
\usepackage[utf8]{inputenc}
\usepackage[english]{babel}
\usepackage{amsfonts}
\usepackage{graphicx}
\usepackage{xcolor}
\usepackage[left=2cm,right=2cm,top=2cm,bottom=2cm]{geometry}
\usepackage{cprotect}
\usepackage[normalem]{ulem}
\usepackage{hyperref}
\usepackage{orcidlink}
\hypersetup{
    colorlinks=true,
    linkcolor=blue,
    filecolor=magenta,      
    urlcolor=cyan,
    pdfauthor={Ond\v{r}ej Zelenka, Bernd Br\"{u}gmann, and Frank Ohme},
    pdftitle={Convolutional Neural Networks for signal detection in real LIGO data}
}

\usepackage[acronym, nomain]{glossaries}
\glsdisablehyper
\setacronymstyle{long-short}
\newacronym[longplural={power spectral densities}]{psd}{PSD}{power spectral density}
\newacronym{gw}{GW}{gravitational wave}
\newacronym{bbh}{BBH}{binary black hole}
\newacronym{mlgwsc1}{MLGWSC-1}{Machine Learning Gravitational-Wave Search Challenge}
\newacronym{far}{FAR}{false-alarm rate}
\newacronym{cnn}{CNN}{convolutional neural network}
\newacronym{usr}{USR}{unbounded softmax replacement}
\newacronym{ligo}{LIGO}{Laser Interferometer Gravitational Observatory}
\newacronym{dain}{DAIN}{deep adaptive input normalization}
\newacronym{ml}{ML}{machine learning}
\newacronym{snr}{SNR}{signal-to-noise ratio}

\def\aus{\textunderscore}
\def\Msun{{M_\odot}}

\begin{document}

\title{Convolutional Neural Networks for signal detection in real LIGO data}
\date{February 12, 2024}

\author{Ond\v{r}ej Zelenka\,\orcidlink{0000-0003-3639-1587}}
\email{ondrej.zelenka@asu.cas.cz}
\affiliation{Friedrich-Schiller-Universit\"{a}t Jena, D-07743 Jena, Germany}
\affiliation{Michael Stifel Center Jena, D-07743 Jena, Germany}
\affiliation{Astronomical Institute of the Czech Academy of Sciences, Bo\v{c}n\'{i} II 1401/1a, CZ-141 00 Prague, Czech Republic}
\author{Bernd Br\"{u}gmann\,\orcidlink{0000-0003-4623-0525}}
\affiliation{Friedrich-Schiller-Universit\"{a}t Jena, D-07743 Jena, Germany}
\affiliation{Michael Stifel Center Jena, D-07743 Jena, Germany}
\author{Frank Ohme\,\orcidlink{0000-0003-0493-5607}}
\affiliation{Max-Planck-Institut f\"{u}r Gravitationsphysik, Albert-Einstein-Institut, D-30167 Hannover, Germany}
\affiliation{Leibniz Universit\"{a}t Hannover, D-30167 Hannover, Germany}

\begin{abstract}
Searching the data of gravitational-wave detectors for signals from compact binary mergers is a computationally demanding task. Recently, machine learning algorithms have been proposed to address current and future challenges. However, the results of these publications often differ greatly due to differing choices in the evaluation procedure. The Machine Learning Gravitational-Wave Search Challenge was organized to resolve these issues and produce a unified framework for machine-learning search evaluation. Six teams submitted contributions, four of which are based on machine learning methods and two are state-of-the-art production analyses. This paper describes the submission from the team TPI FSU Jena and its updated variant. We also apply our algorithm to real O3b data and recover the relevant events of the GWTC-3 catalog.
\end{abstract}

\maketitle

\section{Introduction}

One of the most powerful known sources of \glspl{gw} is a compact binary coalescence: the final stage of a binary composed of compact objects, such as a black hole or neutron star. Analyzing the signal from such an event allows us to constrain the source parameters, such as component masses, which greatly contributes to the study of the black hole population in the universe, and the mechanism by which supermassive black holes are formed~\cite{LIGOScientific:2018jsj, KAGRA:2021duu}. For this reason, \gls{gw} observations are crucial in expanding our understanding of the universe.

Most contemporary detection pipelines are based on \textit{matched filtering}~\cite{LIGOScientific:2021djp} and use a \textit{template bank} of expected waveforms. These pipelines are highly sensitive to signals covered by the template bank, but less sensitive to others. \textit{Loosely modeled searches} are a complementary approach: they do not require the advance knowledge of waveforms to be searched for, but they are less sensitive to compact-binary mergers than matched-filter searches.

With the broadening of the sensitive frequency range of detectors, it becomes necessary to increase the density of template banks. In addition, expanding the parameter-space of interest typically requires more templates to cover the signal manifold. This causes a steep rise in the size of template banks and therefore computational time of matched-filtering based algorithms. In particular, this is an issue when incorporating effects such as eccentricity~\cite{Nitz:2019spj}, precession~\cite{Harry:2016ijz, Schmidt:2023gzj}, or higher modes~\cite{Schmidt:2023gzj, Harry:2017weg}.

Moreover, matched-filter searches are optimal for an idealized Gaussian noise distribution. However, actual detector data deviate from this assumption~\cite{LIGONoiseGuide}. While measures are taken to reduce the effect of this deviation, there are still optimizations to be made. These are some of the driving forces behind the search for new, more efficient methods to complement the matched-filter based analyses.

A rather new development is using \gls{ml} methods in \gls{gw} astronomy. This was started by two pioneering papers on the topic of \gls{gw} detection~\cite{George2018, Gabbard2018}. Their approach consisted of applying convolutional neural networks to recognize whether individual 1 second-long whitened samples of Gaussian noise contain a \gls{bbh} \gls{gw} signal. In another direction, applications in parameter estimation~\cite{Chua:2019wwt, Green:2020hst}, denoising~\cite{Bacon:2022lsm, Shen:2019vep}, fast waveform generation~\cite{Schmidt:2020yuu}, and more~\cite{Cuoco:2020ogp} have also been published; we, however, remain focused on the detection problem in this article.

In the recent years, a multitude of new results have been achieved on this topic~\cite{Schaefer2022_1, Schaefer2022_2, Schaefer2020, Cuoco:2020ogp}. However, due to differing choices in generation of test data, results in the literature are difficult to compare to each other. To resolve this issue, the \Gls{mlgwsc1}~\cite{Schaefer2022_3, mdc_repository} has been organized. From 12 October 2021 until 14 April 2022, multiple teams developed \gls{ml} based algorithms for detection of \gls{gw} signals originating in \gls{bbh} mergers in month-long streams of data from the two US-based \gls{ligo} detectors~\cite{Aasi2015}. The final test data were unknown to participants but followed a known distribution, and no scoreboard was kept during the challenge. Eventually, 4 \gls{ml} based submissions were received, as well as 2 conventional algorithms to provide a baseline. Their performance has been evaluated in detail and effects responsible for differing performance of submissions have been isolated.

We have authored one of the challenge submissions, titled ``TPI FSU Jena''. On test datasets following a simplified Gaussian noise distribution, our search was the top \gls{ml} submission and performed close to the matched-filter baseline, a similar submission being a close second. In addition, it had a comparatively short runtime. However, on test data generated using \gls{ligo} open data~\cite{O3_opendata}, non-Gaussian noise artifacts polluted the search results to a large degree.

In this work, we first briefly describe the \Gls{mlgwsc1}, our submission, and choices made during its development. Following that, we describe the steps taken to further optimize the contribution after the end of the challenge, which greatly improve its performance when non-Gaussian noise transients are present in the data. Finally, we demonstrate the power of the developed searches by applying them to open data from the second half of the third observing run and recovering the GWTC-3 catalog events lying in the relevant portion of the source parameter space.

\section{MLGWSC-1}

\subsection{Test data}\label{ssec:test_data}

The test data consist of 2 strains from the \gls{ligo} Hanford and Livingston detectors. The script used to generate them was available to participants of the challenge with the option to specify its seed. For the final evaluation, a challenge dataset in the length of one month was generated after the challenge deadline using a previously unknown seed~\cite{Schaefer2022_3}.

The test data exist in 4 levels named \textit{datasets} of progressively increasing difficulty. The first three use background noise generated by a colored Gaussian model, while the fourth uses real noise from the O3a observing run~\cite{O3_opendata}. The injection complexity is also increasing, from non-spinning, dominant mode only, to precessing waveforms with generic misaligned spins and multiple higher modes.

Test data are generated using the script \verb|generate_data.py| supplied by the \Gls{mlgwsc1}~\cite{mdc_repository}, which creates the background noise, generates waveforms and injects them into the noise, forming the foreground. Both the background and the foreground are stored in HDF5 files~\cite{hdf5}, each containing groups titled L1 and H1 for the Livingston and Hanford detectors, with the full length of the strain split into multiple segments labeled by their GPS start time. These segments are generated independently of each other. All time series are sampled at a rate of $2048~\mathrm{Hz}$. A low frequency cutoff of $15~\mathrm{Hz}$ is applied to the background noise to allow for reduction in data size of the real detector noise to be downloaded.

The injection parameters are generated by the astrophysical distribution for all angular parameters and the distance is specified by drawing the squared chirp distance uniformly from the interval $d_c^2\in\left[130^2\,\mathrm{Mpc}^2,\, 350^2\,\mathrm{Mpc}^2\right]$. They are placed at random intervals between $24~\mathrm{s}$ and $30~\mathrm{s}$ between merger times. The waveforms are generated using the \verb|IMRPhenomXPHM|~\cite{Pratten:2020ceb} phenomenological model, capable of accurate modeling of precession and higher modes. They are then projected on the corresponding detectors and injected into the background data to produce the foreground.

In the first dataset, Gaussian noise is generated, from the \verb|aLIGOZeroDetHighPower| \acrshort{psd}~\cite{lalsuite, LIGO_document:2009psd} (see Sec.~\ref{ssec:data_processing}). Component masses $m_1,\, m_2 \in \left[10M_\odot,\, 50M_\odot\right]$ are drawn from a uniform distribution, all 6 spin components are set to zero, only the dominant $2, \pm 2$ modes are used, and the low frequency cutoff is chosen to be $20~\mathrm{Hz}$.

In the second dataset, Gaussian noise is generated using an unknown \acrshort{psd}. From a set of 20 \acrshortpl{psd} derived from the O3a observing run data~\cite{O3_opendata}, for each detector, one is randomly chosen and used to generate the noise in all segments. Component masses $m_1,\, m_2 \in\left[7M_\odot,\, 50M_\odot\right]$ are drawn from a uniform distribution, and both spins are aligned with the orbital angular momentum with magnitudes uniformly drawn from a uniform distribution on $\left[-0.99,\, 0.99\right]$.

In the third dataset, noise is generated in a similar manner to the second dataset. However, a new \acrshort{psd} is chosen (from the same set) for each segment. The distribution of component masses is the same as in the second dataset. In contrast, the spins are no longer aligned; their magnitude is uniform from 0 to 0.99, and their direction is isotropically distributed. All higher modes available to the \verb|IMRPhenomXPHM| approximant are used.

In the fourth dataset, real \gls{ligo} noise is used. A real noise file in the extent of approximately 3 months has been prepared by the \Gls{mlgwsc1} team, the data generation script randomly chooses segments from it to comprise the dataset background, and the L1 stream is time-shifted with respect to H1 by a random amount in order to introduce different noise realizations. The injections are generated in a manner identical to the third dataset.

\subsection{Evaluation procedure}

The evaluation is done in a similar manner to \cite{Schaefer2022_1, Schaefer2022_2}. The submitted algorithms are applied to background data without any injections as well as to data with \gls{bbh} injections to determine the relationship between their \gls{far} and sensitive distance.

Each submitted algorithm is required to take a file in the format described in Sec.~\ref{ssec:test_data} as input and produce a file containing identified candidate events as output. It must be an HDF5 file containing 3 datasets referring to the GPS time of the events, the ranking statistics and the tolerance for error in the time.

The evaluation is performed by the \verb|evaluate.py| script supplied by the \Gls{mlgwsc1}~\cite{mdc_repository}. It requires the outputs of the submission algorithm on both the foreground and the background files as input, identifies true positives and determines the \gls{far} at varying detection thresholds. The relationship between the \gls{far} and the sensitive distance is in principle similar to the receiver operating characteristic, which is the relationship between the percentage of false positives and true positives as one varies the threshold for identification of a positive.

To obtain the sensitivity curve of an algorithm based on the identified background and foreground events, we first count the number of background events with a ranking statistic greater than the threshold. Dividing by the total duration of the background data analyzed (in this case $30~\mathrm{days} = 2592000~\mathrm{s}$), we find the \gls{far}. The sensitive volume of the search at $\mathrm{FAR}=\mathcal{F}$ can be calculated by~\cite{Usman2016}
\begin{equation}
V\left(\mathcal{F}\right) = \int \int \epsilon\left(\mathcal{F};\, \mathbf{x},\, \mathbf{\Lambda}\right)\phi\left(\mathbf{x},\, \mathbf{\Lambda}\right)\,\mathrm{d}\mathbf{x}\mathrm{d}\mathbf{\Lambda} ~,
\end{equation}
where $\mathbf{x}$ are an injection's spatial coordinates, $\mathbf{\Lambda}$ the other injection parameters, $\epsilon\left(\mathcal{F};\, \mathbf{x},\, \mathbf{\Lambda}\right)$ is the efficiency of the search, and $\phi\left(\mathbf{x},\, \mathbf{\Lambda}\right)$ is the injection parameter distribution. If we denote $N_{I,\mathcal{F}}$ the number of found injections at a \gls{far} $=\mathcal{F}$ and $\mathcal{M}_{c,i}$, $i=1,\dots,N_{I, \mathcal{F}}$ the chirp masses of the found injections, the expression simplifies to~\cite{Schaefer2022_3}
\begin{equation}
V\left(\mathcal{F}\right) \approx \frac{V\left(d_\mathrm{max}\right)}{N_I}\sum_{i=1}^{N_{I,\mathcal{F}}} \left(\frac{\mathcal{M}_{c,i}}{\mathcal{M}_{c,\mathrm{max}}}\right)^{5/2} ~,
\end{equation}
where $N_I$ is the total number of injections and $\mathcal{M}_{c,\mathrm{max}}$ is the upper limit of injected chirp masses.

We then call the graph of the sensitive volume $V\left(\mathcal{F}\right)$ as a function of the \gls{far} the algorithm's \textit{sensitivity curve} and these are the main criterion for the challenge evaluation.

The runtimes of submitted algorithms were also measured and are available in the challenge paper~\cite{Schaefer2022_3}. All submitted algorithms are evaluated on standardized hardware, provided by the challenge organizers. The hardware consists of a total of 8 Intel Xeon Silver 4215 cores at $2.5~\mathrm{GHz}$, $192~\mathrm{GB}$ of RAM, and 8 nVidia RTX 2070 GPUs with CUDA support, $8~\mathrm{GB}$ of VRAM each.

\section{Experimental setup}\label{sec:experimental_setup}

\subsection{Data processing}\label{ssec:data_processing}

As described in~\cite{LIGONoiseGuide}, the standard noise model in \gls{ligo} detectors is correlated in the time domain. However, using the Fourier transform, in the frequency domain the noise is uncorrelated and described by a Gaussian distribution with zero mean and a frequency-dependent variance called the \gls{psd} and denoted $S_n\left(f\right)$.

Let us use $\tilde{\mathbf{d}}$ to denote the Fourier transform of a time series $\mathbf{d}$. Following~\cite{LIGONoiseGuide}, the transformation
\begin{equation}
\mathbf{d} \mapsto \mathbf{d}_w,\quad \tilde{d}_w\left(f\right) = \frac{\tilde{d}\left(f\right)}{\sqrt{S_n\left(f\right)}}
\end{equation}
yields a time series with a flat \gls{psd}, corresponding to white noise. This process is called \textit{whitening} and is a common method in \gls{gw} data analysis. The \glspl{psd} in \gls{gw} detectors rise steeply towards both low and high frequencies and the signals are dominated by strong noise at frequencies outside the most sensitive band of the detectors.

Following~\cite{Schaefer2022_1}, we feed whitened data to the \gls{ml} model. When applying to test data, the algorithm first estimates the \gls{psd} of the time series in question using Welch's method~\cite{Welch:1967} with a segment duration of $0.5~\mathrm{s}$, then symmetrically truncates the time-domain response of the $S_n\left(f\right)^{-1/2}$ whitening filter to a width of $0.25~\mathrm{s}$, and uses this \gls{psd} to whiten the entire time series. This is done for each segment in the input data separately, as well as for each detector.

As the noise in \gls{ligo} detectors is not stationary over timescales on the order of days, one must account for the \gls{psd} drift. This is addressed by slicing the data into chunks shorter than the \gls{psd}-drift timescale in the test data generation process \cite{mdc_repository}.

\subsection{Training and validation data}

In the training and validation datasets, the noise is taken from the real noise file provided by the \Gls{mlgwsc1}. A segment from the file is chosen at random, its \gls{psd} is estimated and used to whiten the entire segment, and the whitened segment is sliced into 1-second samples. While the noise generation loop is running, these slices are used sequentially, and once all have been used, a new segment is whitened and sliced in the same manner. The \gls{psd} is retained through the processing of the entire segment for whitening of waveform injections.

To generate the waveform injections, we apply the Python package PyCBC~\cite{pycbc_zenodo}. The distributions of individual parameters are summarized in Tab.~\ref{tab:parameters}, they follow the distribution used in test datasets 3 and 4 (see Sec.~\ref{ssec:test_data}) with exceptions, which we describe in the following paragraphs. A limited number of noise samples (given for each experiment in Sec.~\ref{sec:results}) are injected with a waveform and assigned the label $\left(1,\, 0\right)$, the remaining ones remain pure noise and are assigned the label $\left(0,\, 1\right)$. However, the waveforms are normalized to a network optimal \gls{snr} $\rho_\mathrm{net} = 1$ during the data generation procedure and only injected at a randomly generated $\rho_\mathrm{net} \in \left[7,\,20\right]$ at each training epoch. Due to the \gls{snr} normalization, the luminosity distance is irrelevant and a fiducial 1 Mpc value (the PyCBC default) is passed to the approximant.

For consistency with the experiments of~\cite{Schaefer2022_1}, we set the lower mass limit to $10\Msun$ instead of $7\Msun$ used to generate test datasets 2-4. An additional training run confirms that including the range $\left[7\Msun,\, 10\Msun\right]$ in the training data does not improve the performance of the search. We suspect this is due to the increased length of waveforms in this region of the parameter space~\cite{Schaefer2022_3}, due to which a part of the waveform's \gls{snr} is outside the network's input window when the merger is aligned.

Furthermore, due to an oversight on our part, the inclination angle does not follow the astrophysical distribution $\cos\iota \in \left[-1,\, 1\right]$. However, this is not expected to pose an issue, as the dominant effect of the inclination angle on the waveforms is a constant rescaling~\cite{Usman:2018imj}, which is lost as we normalize the waveforms to a fixed network \gls{snr}. A rerun of the code for the \Gls{mlgwsc1} submission with the astrophysical distribution confirms that the results are indistinguishable.

Both the training and validation data are generated by following the steps below:
\begin{enumerate}
\item Get noise:
	\begin{enumerate}
	\item get next slice from current segment
	\item if segment finished, choose a new one at random, whiten it, slice it, and take its first slice
	\end{enumerate}
\item If applicable, generate waveform:
	\begin{enumerate}
	\item set up parameters (see Tab.~\ref{tab:parameters})
	\item generate waveform
	\item crop so that merger is within the given interval, append zeros
	\item whiten using the \gls{psd} of the corresponding noise segment
	\item normalize to optimal $\rho_\mathrm{net} = 1$
	\end{enumerate}
\item Store noise and waveform separately. At each training epoch, inject at a newly generated optimal \gls{snr}.
\end{enumerate}

\begin{table}
\begin{tabular}{lr}
\hline\hline
Parameter\quad & Uniform distribution \\
\hline
Approximant & \verb|IMRPhenomXPHM| \\
Component masses\hspace{3em} & $m_1 \geq m_2 \in \left[10\Msun,\, 50\Msun\right]$ \\
Spin magnitudes & $\left|\boldsymbol\chi_1\right|,\, \left|\boldsymbol\chi_2\right| \in \left[0,\, 0.99\right]$ \\
Spin directions & isotropic \\
Coalescence phase & $\Phi_0 \in \left[0,\, 2\pi\right)$ \\
Inclination angle & $\iota \in \left[0,\, 2\pi\right]$ \\
Declination & $\sin\theta \in \left[-1,\, 1\right]$ \\
Right ascension & $\varphi \in \left[-\pi,\, \pi\right)$ \\
Polarization angle & $\Psi \in \left[0,\, 2\pi\right)$ \\
Sampling rate & $2048~\mathrm{Hz}$ \\
Low frequency cutoff & $20~\mathrm{Hz}$ \\
\hline\hline
\end{tabular}
\caption{Distributions from which waveform injection parameters are drawn. Intervals refer to a uniform distribution.}\label{tab:parameters}
\end{table}

\subsection{Test data}

Test data meant for evaluation before submitting are generated using the program \verb|generate_data.py| supplied by the \Gls{mlgwsc1}~\cite{mdc_repository}. For final testing, all 4 datasets are generated with the length of 2592000~seconds = 30~days, and the seed is set to 4261537. Dataset 4 with this seed is used to generate Fig.~\ref{fig:tpi_e_sel} and to optimize the updated submission in Sec.~\ref{ssec:correction}.

The seeds used for generating the challenge datasets to evaluate the submitted algorithms to the \Gls{mlgwsc1} and to plot Fig.~\ref{fig:mdc_sens} are given in Sec.~\ref{ssec:test_data}.

\subsection{Machine Learning}

The \Gls{mlgwsc1} is aimed at evaluating the performance of \gls{ml} algorithms. In its simplest form, this corresponds to a model with an arbitrary number of free parameters whose error is being optimized over a large dataset. This is frequently done using gradient-descent based optimizers and their stochastic varieties, which approximate the gradients on small batches of the dataset in their successive iterations. The error function being optimized here is a modification of the binary cross entropy loss~\cite{Schaefer2022_1}
\begin{equation}
\mathcal{C}\left(\bar{\mathbf{Y}},\, \mathbf{Y}\right) = -\frac{1}{m}\sum_{i=1}^m \sum_{j=1}^n Y_{ij}\log\left(\left(1 - \varepsilon\right) \bar{Y}_{ij} + \varepsilon\right) ~,
\end{equation}
designed to remove divergences when an element of $\bar{\mathbf{Y}}$ is zero using the regularization parameter $0<\varepsilon \ll 1$.

Neural networks are a class of \gls{ml} models built of artificial neurons, these are functions defined as
\begin{subequations}
\begin{align}
f: \quad &\mathbb{R}^n \to \mathbb{R} ~,\\
&\mathbf{x} \mapsto \sigma\left(\sum_{i=1}^n w_ix_i + b\right) ~.
\end{align}
\end{subequations}
The parameters $w_i$ and $b$ are called the weights and bias, respectively, and are optimized through the training process. The function $\sigma$ is called an \textit{activation function}, a popular choice we use here is the Exponential Linear Unit~\cite{Clevert:2015qvd} with $\alpha=1$
\begin{equation}
\mathrm{ELU}\left(z\right) = \begin{cases} \alpha\left(\exp\left(z\right) - 1\right) & \text{if } z<0~,\\ z & \text{if } z\geq 0~.\end{cases}
\end{equation}

A feed-forward neural network is organized in layers of independent neurons, each of which feeds its output into neurons of the following layer. They can be fully connected, i.e. the input of each neuron consists of the outputs of all neurons in the previous layer, also called dense layers. In this paper, we also make use of convolutional layers, whose structure corresponds to a set of filters sliding over a multichannel input. This reduces the number of independent connections and thus weights in the network.

Further components are max pool layers, which act as a downsampling operation, and dropout layers, which improve the training convergence through a type of noise injection. For an introduction to \gls{ml} and neural networks, we refer the reader to~\cite{Mehta:2018dln, Goodfellow:2016aaa}.

\subsection{Model architecture}

In this work, we use a simple \gls{cnn} design, which is an extension of the architecture used in \cite{Schaefer2022_1}. For a simple implementation of the method used there, we first define a \gls{cnn} called the \textit{base network}, which does not have a final activation. Its architecture is shown in Table \ref{tab:base_net}.

Unlike coincident searches such as the CNN-Coinc submission~\cite{Schaefer2022_2} to the \gls{mlgwsc1}, wherein the streams from each detector are analyzed separately and combined using a probability-based formula, we employ a coherent approach. The network accepts a two-channel input to carry data from two detector streams.

\begin{table}
\begin{tabular*}{0.75\linewidth}{@{\extracolsep{\fill}} lccc}
\hline\hline
layer & KS & shape & activation \\
\hline
input & & $2\times 2048$ & \\
(batch norm) & & $2\times 2048$ & \\
convolution & 33 & $16\times 2016$ & ELU \\
convolution & 32 & $16\times 1985$ & ELU \\
convolution & 17 & $16\times 1969$ & ELU \\
convolution & 16 & $16\times 1954$ & ELU \\
max-pool & 4 & $16\times 488$ & \\
convolution & 17 & $16\times 472$ & ELU \\
convolution & 16 & $32\times 457$ & ELU \\
convolution & 9 & $32\times 449$ & ELU \\
convolution & 8 & $32\times 442$ & ELU \\
max-pool & 3 & $32\times 147$ & \\
convolution & 9 & $32\times 139$ & ELU \\
convolution & 8 & $64\times 132$ & ELU \\
convolution & 9 & $64\times 124$ & ELU \\
convolution & 8 & $64\times 117$ & ELU \\
max-pool & 2 & $64\times 58$ & \\
flatten & & $3712$ & \\
dense & & $128$ & ELU \\
dropout & & $128$ & \\
dense & & $128$ & ELU \\
dropout & & $128$ & \\
dense & & $2$ & \\
\hline\hline
\end{tabular*}
\caption{Architecture of the base network. It accepts an input with 2 channels corresponding to 2 detector streams and possesses 635318 trainable weights. ``KS'' refers to kernel size, and ``shape'' is the output shape of the corresponding layer. The batch normalization layer is only used in the original submission to the \Gls{mlgwsc1} but not in the improved searches.}\label{tab:base_net}
\end{table}

The base network produces 2 outputs, which we denote $x_0$, $x_1$. Following the method of \cite{Schaefer2022_1}, for training we append a Softmax layer
\begin{equation}\label{eq:softmax}
y_i = \mathrm{Softmax}\left(\mathbf{x}\right)_i = \frac{\exp\left(x_i\right)}{\sum_j \exp\left(x_j\right)} ~,
\end{equation}
which maps its inputs to a set of positive numbers which sum up to one. The purpose of this activation is to represent probabilities of different classes in classification problems, and in this case we wish the output $y_0$ to represent the probability that the input sample contains an astrophysical \gls{gw} signal.

The networks are trained using the stochastic Adam optimizer \cite{adam} with a learning rate of $\gamma = 4\cdot 10^{-6}$, and the other parameters set to their defaults in PyTorch \cite{pytorch_software} ($\beta_1=0.9$, $\beta_2=0.999$, $\epsilon=10^{-8}$), for a total of 250 epochs.

When testing in the same manner, however, a numerical issue arises. In single-precision floating point arithmetic using PyTorch, $y_0$, which we would like to use as the ranking statistic of the resulting search, rounds up to 1 when $x_0-x_1 \gtrsim 16$, which is well in the range of values encountered by the search. To resolve this, we rewrite Eq.~\eqref{eq:softmax} for $i=0$ as
\begin{equation}\label{eq:usr}
y_0 = \frac{1}{1+\exp\left(x_1-x_0\right)} = \frac{1}{1+\exp\left(-\Delta x\right)} ~.
\end{equation}
We see that $y_0$ is a purely growing function of $\Delta x = x_0 - x_1$, which is therefore an equivalent ranking statistic, without suffering from the same numerical issue. Therefore, $\Delta x$ is used as the ranking statistic in the search. This technique is called the \gls{usr}. For more detailed information see~\cite{Schaefer2022_1}.

\begin{figure}[t]
  \includegraphics[width=\linewidth]{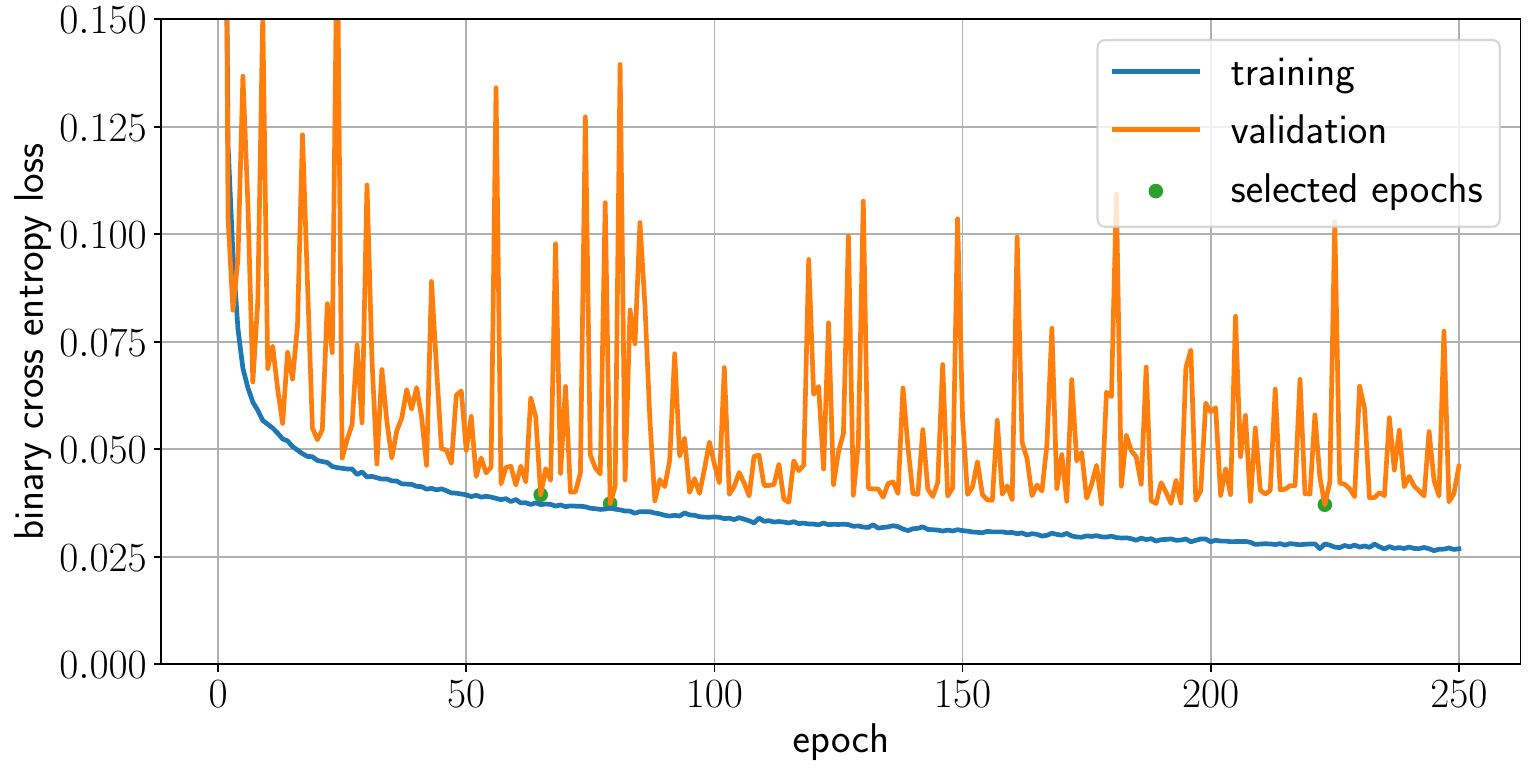}
  \caption{Evolution of the training and validation loss values throughout the training of the \Gls{mlgwsc1} submission.}\label{fig:tpi_losses}
\end{figure}
\begin{figure}
  \includegraphics[width=\linewidth]{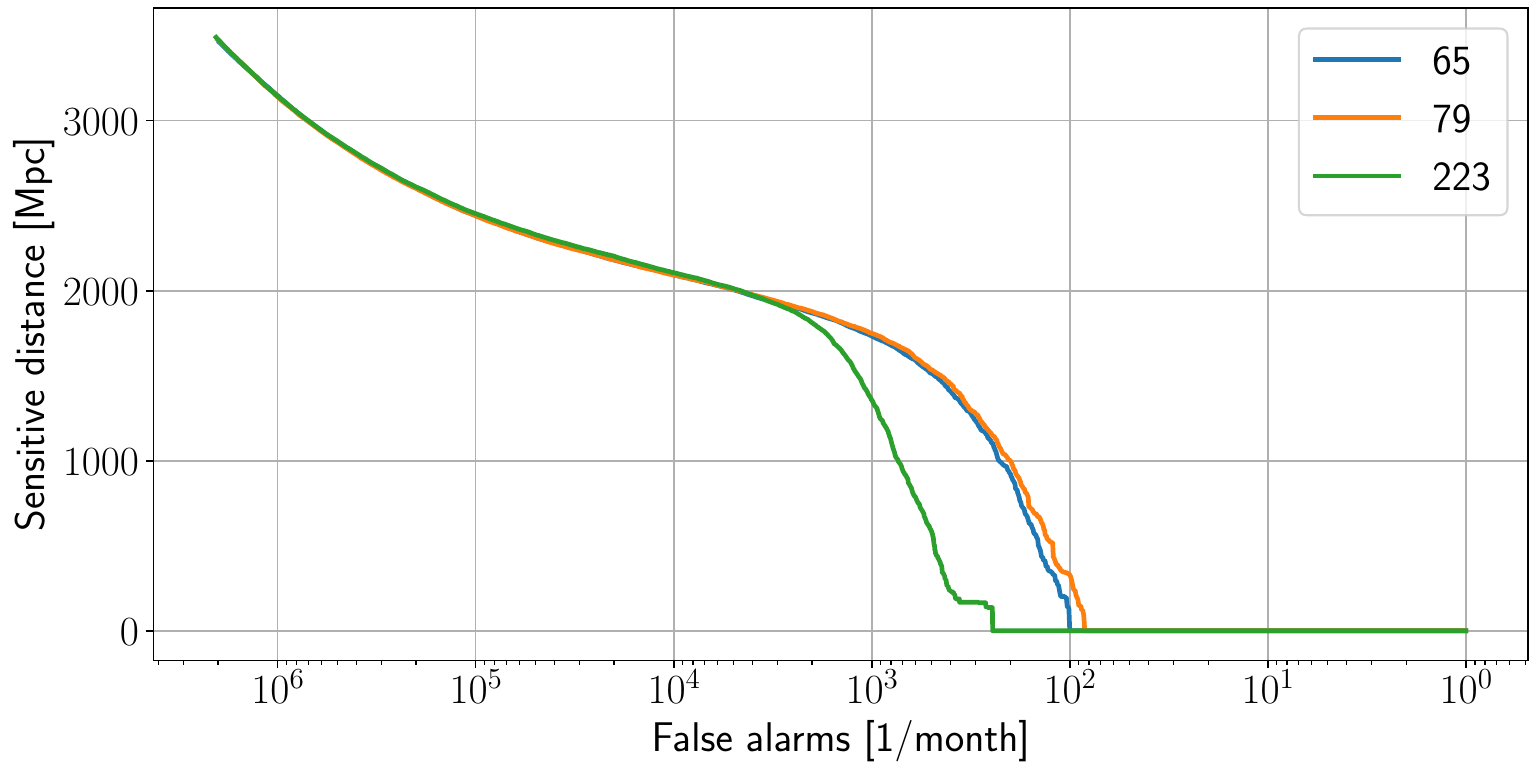}
  \caption{Sensitivity curves of the network at 3 minima of the validation loss highlighted in Fig.~\ref{fig:tpi_losses} used to select the final network state for the submission.}\label{fig:tpi_e_sel}
\end{figure}

The \gls{cnn} is only part of the detection algorithm following~\cite{Schaefer2022_1}, as it only accepts simple 1-second-long slices. The full algorithm consists of feeding overlapping slices of the test data to the network, applying a threshold, and clustering the results into candidate detections. First, the entire segment is whitened using the method described in Sec.~\ref{ssec:data_processing}.

Then, the segment is sliced into 1-second long samples with an offset of 0.1 seconds, which are fed to the network and the $\Delta x$ outputs recorded. Because the networks are trained on injections with merger time 0.6 to 0.8 seconds after the sample start time, each slice is associated with the time 0.6 seconds after the start time of the slice, in order to compensate for this alignment.

A threshold is applied to the network outputs and those which exceed it are clustered by time, with a minimal separation of 0.35 seconds between clusters. Each of these clusters is then considered a candidate event to be saved in the output file. As the ranking statistic, the maximum of the network outputs in the cluster is used, and the time corresponding to the maximum is used as the time of the candidate event. For all output events, the value of 0.2 seconds is chosen as the time uncertainty, to match the size of the merger alignment interval in the training data.

\section{Results}\label{sec:results}
\subsection{MLGWSC-1 submission}\label{ssec:mdc_submission}

\begin{figure}
  \includegraphics[width=\linewidth]{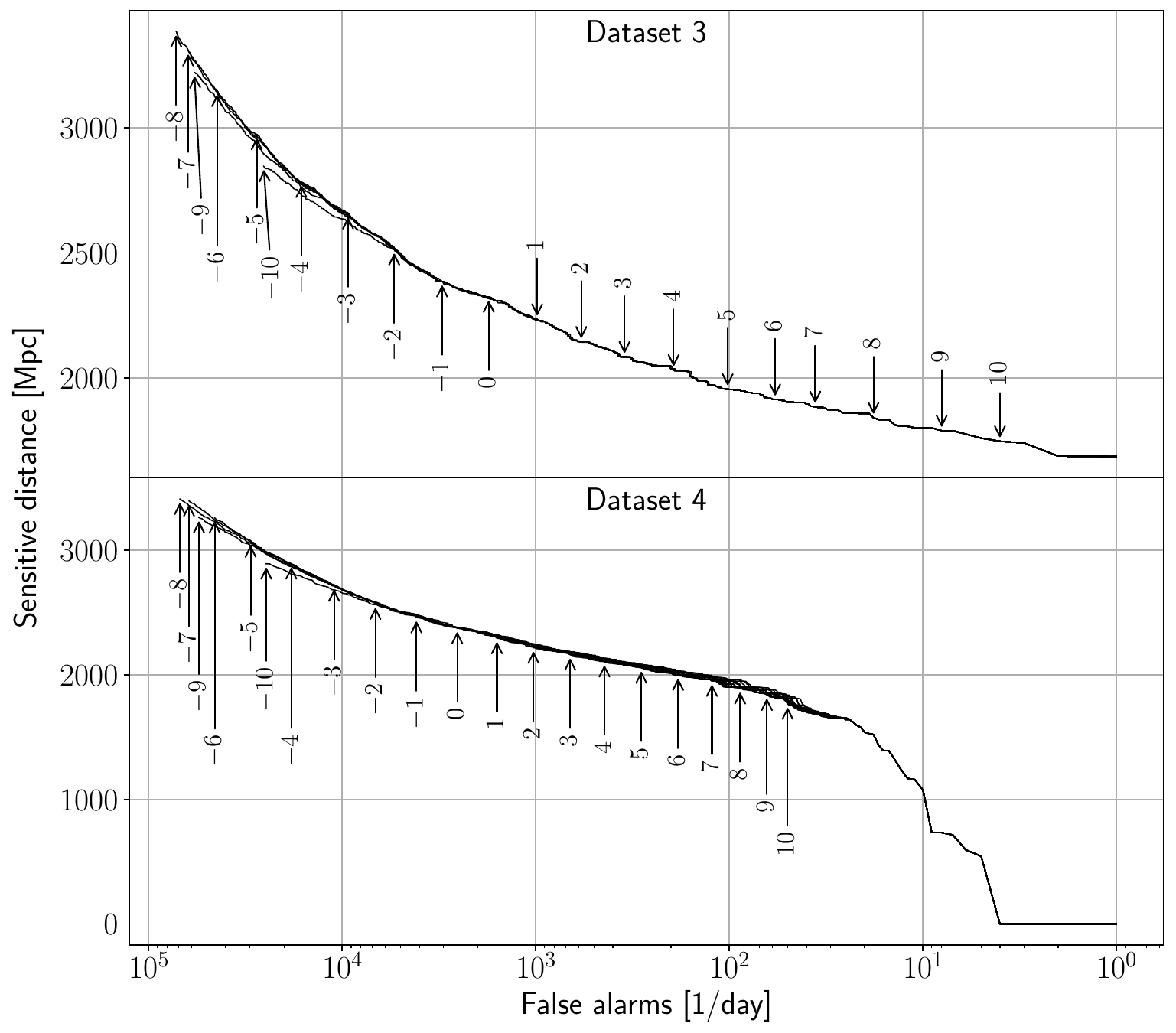}
  \caption{Sensitivity curves of the submitted trained network using multiple $\Delta x$ thresholds to determine a suitable value. The datasets are generated by the script provided by the~\Gls{mlgwsc1} in the length of 1 day (86400 seconds), difficulty specified as datasets 3 and 4 in the top and bottom panel, respectively. Due to a large overlap between the sensitivity curves, rather than color-coding, their left ends are annotated with the threshold value. All curves reach the same point at the right end $\mathcal{F} = 1\,\mathrm{day}^{-1}$.}\label{fig:mdc_thresholds}
\end{figure}

For the submission, we choose the training dataset to contain 500000 pure noise samples and 500000 noise+waveform samples, the validation dataset to contain 100000 pure noise samples and 100000 noise+waveform samples, and the sliced real noise is used. The training and validation losses are monitored during the training, and their evolution is shown in Fig.~\ref{fig:tpi_losses}.

Out of the local minima of the validation loss, the global minimum as well as two earlier local minima are chosen and further tested by applying to the test datasets 3 and 4, the result for dataset 4 is shown in Fig.~\ref{fig:tpi_e_sel}. The results on dataset 3 were virtually indistinguishable, for better performance on dataset 4 we choose the network state at epoch 79 for the submission to the \Gls{mlgwsc1}.

\begin{figure*}[t]
  \includegraphics[width=\textwidth]{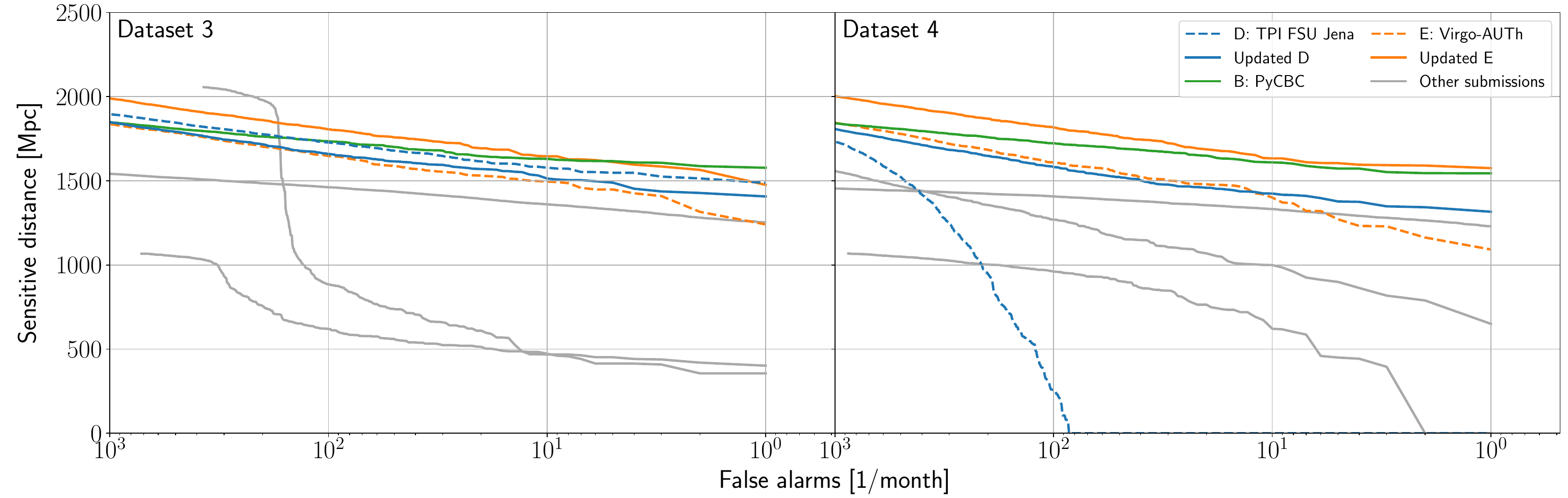}
  \caption{Sensitivity curves of 3 selected submissions, along with updated versions of 2 of them, on datasets 3 and 4 of the \Gls{mlgwsc1}. Each panel contains the performance of the submissions on one test dataset. Dashed lines mark conventional analyses and solid lines mark \gls{ml}-based search algorithms. In case of the TPI FSU Jena and Virgo-AUTh teams, the dotted lines mark the original submissions, while the solid lines mark the updated algorithms. The remaining submissions are shown in gray for illustration of overall challenge results.}\label{fig:mdc_sens}
\end{figure*}

It is necessary to set one more parameter: the first detection threshold, applied before clustering. Applying the trained algorithm to a shorter dataset (length of 1 day), the resulting sensitivities are displayed in Fig.~\ref{fig:mdc_thresholds}. We choose the value of -8 for the threshold applied to the $\Delta x$ ranking statistic, as its performance is indistinguishable from others at lower \glspl{far}, while it reaches up to higher \glspl{far} than others. However, at \gls{far} values relevant to \gls{gw} astronomy, all algorithms perform comparably, therefore this choice does not seem to be relevant and we do not perform the same optimization in further experiments.

\subsection{MLGWSC-1 results}

The \Gls{mlgwsc1} received a total of six contributions, four of which are \gls{ml} based. The remaining two are conventional analyses to provide a baseline; the first is the matched-filtering based PyCBC~\cite{Usman2016}, the other is the loosely modeled search cWB~\cite{Klimenko2016, cwb_repo}.

Rather than an extensive coverage of the \Gls{mlgwsc1} results, which are described in great detail in~\cite{Schaefer2022_3}, this section focuses on a particular issue which occurs when real noise is presented to our algorithm. We would like to specifically bring to the reader's attention the performance of our algorithm (labeled D: TPI FSU Jena) and the algorithm labeled E: Virgo-AUTh, whose sensitivity curves on datasets 3 and 4 are shown in Fig.~\ref{fig:mdc_sens}. Both \gls{ml} submissions are plotted as dashed lines, in addition the PyCBC submission is shown.

Both submissions use a very similar approach. In the final testing, their performances are close to each other with D operating at a slightly higher sensitivity at all \glspl{snr}, this gap widening as we approach $\mathcal{F} = 1~\mathrm{month}^{-1}$, on test dataset 3 (in fact, this holds on all datasets which use Gaussian noise~\cite{Schaefer2022_3}). However, the Virgo-AUTh algorithm retains $\geq 90\%$ of the sensitive distance of the TPI FSU Jena search at $\mathcal{F} \geq 2~\mathrm{month}^{-1}$, and at $\mathcal{F} = 1000~\mathrm{month}^{-1}$ this gap narrows to a separation of roughly $4\%$.

Moving to dataset 4, the performance of the Virgo-AUTh algorithm degrades only mildly. In contrast, the performance of our submission deteriorates much more, losing all sensitivity at $\mathcal{F} < 10^2~\mathrm{month}^{-1}$. This is due to the noise transients which are omnipresent in real detector data, and which are revealed to produce triggers louder than injected waveforms in further analysis.

Finally, the runtimes of our algorithm are consistently lower than those of the other submissions. On average, the Virgo-AUTh search takes $\sim\negmedspace 50\%$ longer to run on the challenge hardware on all 4 test datasets due to the higher complexity of its network architecture. On datasets 2-4 the estimated runtimes of PyCBC are $\sim\negmedspace 40$ times as large. We note that the given PyCBC runtimes are estimations as a different hardware setup is used to run the search.

\subsection{Updated submission}\label{ssec:correction}

Following the Virgo-AUTh team's algorithm through~\cite{Schaefer2022_3, Nousi2022, github_aresgw}, we identify 3 main differences, which we expect to be responsible for the large difference in performance. These are: input normalization, network architecture, and training dataset distribution, and we cover them in detail below.

\subsubsection{Input normalization and network architecture}

While our submission retains the batch normalization layer from~\cite{Schaefer2022_1}, the Virgo-AUTh team has tested multiple input normalization methods, and uses a \gls{dain} layer instead. Inspired by this idea, we attempt to replace the batch normalization layer by a \gls{dain}, as well as simply remove the input normalization altogether. Out of these three options, only the complete removal of input normalization brings a discernible reduction in \glspl{far}.

While our submission makes use of a fairly simple and small \gls{cnn} design, the Virgo-AUTh team's submission uses a larger and more complex ResNet design. We attempt to replace our network with an identical ResNet design and encounter no improvement in sensitivity, along with an increased computational cost. The same experiment with all three input normalization options mentioned above yields similar results. Therefore, we decide to retain the original \gls{cnn} design and merely remove the batch normalization layer for further development.

\subsubsection{Training dataset distribution}\label{sssec:dataset_distro}

In the end, the key issue turns out to be the distribution of the training dataset. We use a 1:1 dataset in terms of the number of pure noise samples to ones with injections in our original submission. Further experiments indicate the optimal ratio to be 1:3 as the network's performance degrades when this ratio is shifted in either direction.

Six training runs are performed using the same optimization procedure as previously. At each epoch, the network's sensitivity is evaluated on test data with real noise, and from each run the state with the highest sensitive distance at $\mathcal{F} = 1\,\mathrm{month}^{-1}$ is chosen and labeled as \verb|R<run number 1-6>/<4-digit epoch number>|. Of the resulting 6 states, we choose \verb|R1/0021| for the final search algorithm as it has the highest sensitivity. Its sensitivity curves are shown in Fig.~\ref{fig:mdc_sens} alongside the curves of all submissions as well as the updated Virgo-AUTh search~\cite{Nousi2022, github_aresgw}.

The sensitivity on datasets using Gaussian noise deteriorates slightly;  this is to be expected as one optimizes for a different noise distribution, rejecting potential glitches in data containing none. At $\mathcal{F}=1~\mathrm{month}^{-1}$, the sensitive distance is reduced by $5.4\%$. In the overall ranking, ours remains the most sensitive of all \gls{ml} submissions on Gaussian noise.

On real noise, the updated submission reaches the highest sensitivity of all \gls{ml} submissions at $\mathcal{F} \lesssim 10~\mathrm{month}^{-1}$ and is narrowly outperformed by Virgo-AUTh at higher \glspl{far}. At $\mathcal{F} = 1~\mathrm{month}^{-1}$, our updated submission has a sensitivite distance of $1316~\mathrm{Mpc}$, and Virgo-AUTh operates at $87\%$ of this value. At the same time, the updated version of their algorithm outperforms ours in both cases.

\subsection{Application to O3b data}\label{ssec:O3b_application}

The O3 \gls{ligo} observing run was split by a commissioning break into two phases, O3a and O3b~\cite{O3_opendata, KAGRA:2013rdx}. The first part is used to train the \glspl{cnn} above to recognize \gls{bbh} waveform injections in real \gls{ligo} noise. In this section, we apply the searches developed above to real data recorded by \gls{ligo} through the O3b phase and cross-reference the output with the transients recorded in the GWTC-3 catalog~\cite{LIGOScientific:2021djp}.

To query O3b data, we require a minimum segment length of one minute and the same data quality requirements as the real noise file used in the \Gls{mlgwsc1}, known injections are not removed. This leaves us with a total of 8~228~706 seconds of data in a total of 2377 segments, amounting approximately to 95 days and 6 hours. In comparison, the full O3b observing run was 147 days and 2 hours in length.

We apply all 6 searches trained in Sec.~\ref{sssec:dataset_distro} to these data. The GWTC-3 catalog~\cite{LIGOScientific:2021djp} consists of 35 confident detections and 7 marginal ones. Events lying outside the segments of available data are excluded, leaving us with 31 confident and 4 marginal events to be found. These excluded events are listed in the left column of Tab.~\ref{tab:O3b_omitted}. In addition, we confirm that none of the events contained in available segments take place closer to either end of their respective segments than 46 seconds.

\begin{table}
\begin{tabular*}{\linewidth}{@{\extracolsep{\fill}} lccccccc}\hline\hline
 & & \rotatebox{90}{\small R1/0021\,} & \rotatebox{90}{\small R2/0150\,} & \rotatebox{90}{\small R3/0036\,} & \rotatebox{90}{\small R4/0038\,} & \rotatebox{90}{\small R5/0193\,} & \rotatebox{90}{\small R6/0026\,} \\
Event name & $\rho_\mathrm{MF}$ & \multicolumn{6}{c}{$\mathcal{F}\,\left[\mathrm{month}^{-1}\right]$} \\
\hline
GW200224\aus 222234 & 20.0 & 0.0 & 0.9 & 0.0 & 0.0 & 0.0 & 0.0 \\
GW200311\aus 115853 & 17.8 & 0.0 & 0.9 & 0.6 & 1.3 & 0.0 & 6.0 \\
GW200225\aus 060421 & 12.5 & 0.0 & 1.6 & 0.0 & 0.0 & 1.9 & 0.3 \\
GW191215\aus 223052 & 11.2 & 0.0 & 1.9 & 1.3 & 1.3 & 0.0 & 1.3 \\
GW200208\aus 130117 & 10.8 & 19.2 & 2.2 & 3.5 & 3.1 & 1.6 & 10.1 \\
GW200219\aus 094415 & 10.7 & 5.0 & 4.7 & 8.8 & 38.7 & 12.6 & 19.5 \\
GW200209\aus 085452 & 9.6 & 1.3 & 2.8 & 0.9 & 2.8 & 2.2 & 0.3 \\
GW191204\aus 110529 & 8.8 & 1.6 & 3.1 & 0.0 & 3.1 & 5.0 & 3.1 \\
GW200308\aus 173609 & 7.1 & - & - & - & - & - & - \\
\hline
GW191222\aus 033537 & 12.5 & 0.0 & 4.1 & 2.5 & 2.5 & 0.3 & 0.3 \\
GW200128\aus 022011 & 10.6 & 25.5 & 3.1 & 0.0 & 11.7 & 10.4 & 2.2 \\
GW191230\aus 180458 & 10.4 & 6.6 & 149 & 19.5 & 98.9 & 36.9 & 5.0 \\
GW191127\aus 050227 & 9.2 & 38.7 & 2.8 & 4.4 & 18.6 & 3.1 & 6.6 \\
GW200220\aus 124850 & 8.5 & 215 & 517 & 956 & 96.1 & 695 & 375 \\
GW191126\aus 115259 & 8.3 & - & - & - & - & - & - \\
GW200216\aus 220804 & 8.1 & - & 189 & - & - & 841 & - \\
GW191113\aus 071753 & 7.9 & - & 634 & 391 & - & 713 & 647 \\
GW200306\aus 093714 & 7.8 & 485 & 407 & 720 & - & 69.3 & - \\
GW200208\aus 222617 & 7.4 & 38.1 & 6.0 & 19.5 & 55.1 & 159 & 187 \\
GW200322\aus 091133 & 6.0 & 810 & 898 & - & - & - & - \\
\hline
GW191204\aus 171526 & 17.5 & 3.5 & 8.8 & 4.1 & 4.4 & 7.6 & 6.0 \\
GW191109\aus 010717 & 17.3 & 0.0 & 1.9 & 0.9 & 0.6 & 1.3 & 0.9 %\\
%GW191129\aus 134029 & 13.1 & - & - & - & - & - & - \\
%GW200115\aus 042309 & 11.3 & - & - & - & - & - & - \\
%GW200202\aus 154313 & 10.8 & - & - & - & - & - & - \\
%GW200316\aus 215756 & 10.3 & - & - & - & - & - & - \\
%GW191105\aus 143521 & 9.7 & - & - & - & - & 658 & - \\
%GW191219\aus 163120 & 9.1 & - & - & - & - & - & - \\
%GW191103\aus 012549 & 8.9 & - & - & - & - & - & - \\
%GW200210\aus 092254 & 8.4 & - & - & - & - & - & - \\
%GW200220\aus 061928 & 7.2 & - & - & - & - & - & -

\\\hline\hline\end{tabular*}
\cprotect\caption{List of O3b events from the \verb|GWTC-3-confident| catalog~\cite{LIGOScientific:2021djp} and their identification by the 6 final searches. Events which are not recovered by the given search are marked by a hyphen. 13 events are omitted (see Tab.~\ref{tab:O3b_omitted}). The events are grouped into three sections based on their estimated component masses (see text for details).}\label{tab:O3b_results}
\end{table}

A catalog event is marked as found, if the search output contains an event within 0.2 seconds of the time given in the catalog, and it is assigned its corresponding ranking statistic $t$. The remaining catalog events are considered missed, and the remaining events reported by the search are considered false alarms. The catalog event is then considered detected at a \gls{far} of
\begin{equation}
\mathcal{F} = \frac{N_{f>t}}{T} ~,
\end{equation}
where $N_{f>t}$ is the number of false alarms louder than $t$, and $T$ is the total length of the analyzed segments. In addition, if the \gls{far} of an event is at least $1000~\mathrm{month}^{-1}$, it is also considered missed.

\begin{table}
\begin{tabular*}{.75\linewidth}{@{\extracolsep{\fill}} cc}\hline\hline
data quality & missed \\
\hline
GW200302\aus 015811 & GW191129\aus 134029 \\
GW200129\aus 065458 & GW200115\aus 042309 \\
GW200112\aus 155838 & GW200202\aus 154313 \\
GW191216\aus 213338 & GW200316\aus 215756 \\
                    & GW191105\aus 143521 \\
                    & GW191219\aus 163120 \\
                    & GW191103\aus 012549 \\
                    & GW200210\aus 092254 \\
                    & GW200220\aus 061928 
\\\hline\hline\end{tabular*}
\cprotect\caption{List of O3b events omitted from Tab.~\ref{tab:O3b_results}. Events listed in the first column are omitted due to insufficient data quality in either detector, and events listed in the second column are omitted due to being missed completely by all searches. The exception is GW191105\aus 143521, which is recovered at $\mathcal{F} = 658~\mathrm{month}^{-1}$ by the \verb|R5/0193| search and missed by the others.}\label{tab:O3b_omitted}
\end{table}

None of the events marked marginal in the GWTC-3 catalog are found by either of the searches. The resulting \glspl{far} of confident events in the analyzed segments are shown in Tab.~\ref{tab:O3b_results}. The table is split into three sections: in the first, the 90\% credible intervals on both component masses lie fully in the $\left[10\Msun,\, 50\Msun\right]$ range used for training the networks, while in the third, at least one of them lies fully outside $\left[10\Msun,\, 50\Msun\right]$. The remaining cases are contained in the second section. The credible intervals and accompanying \gls{snr} values come from the catalog's parameter estimation pipeline based on Bilby~\cite{Ashton:2018jfp, Romero-Shaw:2020owr} and are supplied by GWOSC~\cite{gwosc_gwtc3_strains}.

Let us comment shortly on the results of Tab.~\ref{tab:O3b_results}. Most importantly, all events in the first section, where the search algorithms are expected to operate at a high sensitivity, are found by all 6 tested networks at a \gls{far} lower than $40\,\mathrm{month}^{-1}$, with the exception of GW200308\textunderscore 173609, which is the second weakest event in the catalog at $\rho_\mathrm{MF} = 7.1$. In the vast majority, the events are detected at $\mathcal{F} < 4\,\mathrm{month}^{-1}$.

In the second and third sections the searches are expected to operate at a reduced sensitivity as the corresponding parameter space is not fully covered in the training dataset. This is confirmed in Tab.~\ref{tab:O3b_results}, however, louder events at $\rho_\mathrm{MF} \gtrsim 9$ and $\rho_\mathrm{MF} \gtrsim 17$ in the second and third section, respectively, are also mostly detected at $\mathcal{F} < 10\,\mathrm{month}^{-1}$ by the \gls{ml}-based searches.

As a final comment, Q-scan spectrograms of the loudest events in the analyzed data seem to be consistent with the false alarms being known types of glitches. Full outputs of all 6 search algorithms as well as spectrograms of the 128 loudest events of each are publicly available in the data release~\cite{ml_gw_repository}.

\section{Conclusion}

We have presented a \acrlong{cnn}-based \acrlong{gw} detection algorithm capable of performing comparably to conventional algorithms in specific settings, and its implementation, submitted to the \gls{mlgwsc1}. While the submission performs well on test data using Gaussian noise, the noise transients present in the data with real noise prove to be too much of a challenge and reduce its sensitivity to zero at relevant \glspl{far}. In the present work, we resolve this issue by a careful optimization of the training parameters and demonstrate that the updated search outperforms all other submissions besides the PyCBC matched-filter search.

At the same time, while each independent run of the updated algorithm converges to a state with high sensitivity of the resulting search, a detailed analysis reveals that the sensitivity is highly non-monotonic during the training~\cite{Zelenka:2023phd}. In addition, Fig.~\ref{fig:tpi_losses} also shows unexpected oscillations in the validation loss. This phenomenon is not yet fully understood and warrants further investigation.

As a final application of the updated search, we analyze open data from the O3b observing run~\cite{O3_opendata} of the \gls{ligo}-Virgo collaboration and cross-reference the results with the corresponding catalog GWTC-3~\cite{LIGOScientific:2021djp}. We demonstrate that in the intended regime of \glspl{bbh} with component masses between $10\Msun$ and $50\Msun$, our searches can confidently detect events with a network \gls{snr} above 8. This is in line with contemporary matched-filter based searches, as the value 8 roughly corresponds to 1 false alarm per month~\cite{Schaefer2022_1}.

\section{Acknowledgments}

O.Z. thanks the Carl Zeiss Foundation for the financial support within the scope of the program line ``Breakthroughs'' and is supported by the fellowship Lumina Quaeruntur No. LQ100032102 of the Czech Academy of Sciences. Further support has been provided by the COST network CA17137 "G2net".

The computational experiments were performed on the ARA cluster at the Friedrich-Schiller-Universit\"{a}t Jena and the Atlas cluster financed by the Gottfried Wilhelm Leibniz Universit\"{a}t Hannover and the Max-Planck-Gesellschaft through the Albert-Einstein-Institut Hannover. We thank the Observational Relativity and Cosmology division for access.

Special thanks go to Marlin Sch\"{a}fer for his great contribution through his work on organizing the \Gls{mlgwsc1} as well as discussions, and to all contributors to the challenge.

This research has made use of data or software obtained from the Gravitational Wave Open Science Center (gwosc.org), a service of the LIGO Scientific Collaboration, the Virgo Collaboration, and KAGRA. This material is based upon work supported by NSF's LIGO Laboratory which is a major facility fully funded by the National Science Foundation, as well as the Science and Technology Facilities Council (STFC) of the United Kingdom, the Max-Planck-Society (MPS), and the State of Niedersachsen/Germany for support of the construction of Advanced LIGO and construction and operation of the GEO600 detector. Additional support for Advanced LIGO was provided by the Australian Research Council. Virgo is funded, through the European Gravitational Observatory (EGO), by the French Centre National de Recherche Scientifique (CNRS), the Italian Istituto Nazionale di Fisica Nucleare (INFN) and the Dutch Nikhef, with contributions by institutions from Belgium, Germany, Greece, Hungary, Ireland, Japan, Monaco, Poland, Portugal, Spain. KAGRA is supported by Ministry of Education, Culture, Sports, Science and Technology (MEXT), Japan Society for the Promotion of Science (JSPS) in Japan; National Research Foundation (NRF) and Ministry of Science and ICT (MSIT) in Korea; Academia Sinica (AS) and National Science and Technology Council (NSTC) in Taiwan.

\bibliographystyle{apsrev4-2}
\bibliography{bibliography}

\appendix
\section{Code}\label{app:code}
The code provided by the \gls{mlgwsc1} organizers is available in~\cite{mdc_repository}. The code used in the experiments is contained in \cite{ml_gw_repository}. The submission to the \Gls{mlgwsc1} as described in Secs.~\ref{sec:experimental_setup}, \ref{ssec:mdc_submission} is stored in the directory \verb|mlgwsc-1|. For the experiments detailed in Secs.~\ref{ssec:correction}, \ref{ssec:O3b_application}, the code is available in the subdirectory \verb|correction| along with additional materials and results.

\end{document}